# MAGNETIC UNIPOLAR FEATURES IN CONDUCTIVITY OF POINT CONTACTS BETWEEN NORMAL AND FERROMAGNETIC D-METALS (CO, NI, FE)

Yu. G. Naidyuk (`naidyuk@ilt.kharkov.ua`), I. K. Yanson, D. L. Bashlakov, V. V. Fisun, O. P. Balkashin, L. Y. Triputen'
*B.Verkin Institute for Low Temperature Physics and Engineering, National Academy of Sciences of Ukraine, 47 Lenin Ave., 61103, Kharkiv, Ukraine*

A. Konovalenko, V. Korenivski
*Nanostructure Physics, Royal Institute of Technology (KTH), Stockholm, Sweden*

R. I. Shekhter
*Department of Physics, Göteborg University, SE–412 96 Göteborg, Sweden*

**Abstract.** In nanocontacts between normal and ferromagnetic metals (N–F) abrupt changes of the order of 1 % are detected in differential resistance, $dV/dI(V)$, versus bias voltage, $V$, on achieving of high current densities, $10^9$ A/cm$^2$. These features in $dV/dI(V)$ are observed when the electron flow is directed from the nonmagnetic metal into the ferromagnet and connected with magnetization excitations in the ferromagnet induced by the current. Applying an external magnetic field leads to a shift of the observed features to higher biasing current, confirming the magnetic nature of the effect. Such effects are observed for the non-ballistic (not spectral) regime of current flow in the nanocontacts. Thus, the current induced magneto-conductance effects in multilayered N–F structures (nanopillars) extensively studied in the recent literature have much more general character and can be stimulated by elastic electron scattering at single N–F interfaces.

## 1. Introduction

Nonlinearity of the current-voltage characteristics (IVC) of point contacts (PC) contains direct information about interaction of the conduction electrons with various quasiparticle excitations in metallic solids (Naidyuk and Yanson, 2005). It turns out that PC spectra (the second derivative of IVC) are proportional to the spectral function of electron-quasiparticle interactions. A necessary condition for obtaining such spectra is a small PC size – the size of the PC $d$ should be smaller than the inelastic electron relaxation length $\Lambda_\varepsilon$. Additionally, to achieve the necessary spectral resolution the measurement temperature should be much lower than the characteristic energy of the relevant quasiparticle excitations. Studies of PC spectra for both usual (normal) and ferromagnetic metals have revealed features of the electron-phonon interaction (EPI) function (Naidyuk and Yanson, 2005; Khotkevich and Yanson, 1995) and the electron-magnon interaction function (Naidyuk and Yanson, 2005).

In the present paper we investigate new features of non-spectral character in $dV/dI(V)$ and $d^2V/dI^2(V)$ of heterocontacts between normal metals (N=Cu, Ag) and ferromagnets (F=Co, Ni, or Fe). These features are observed most distinctly in an external magnetic field and are incompatible with the spectral EPI features. They originate from disturbances of the magnetic order at the ferromagnetic surface, under the influence of extremely high current density of electrons incident from the normal metal. Similar features were observed earlier in lithographically prepared ferromagnetic nanoconstrictions (Ralph, 1994; Myers et al., 1999) or in PCs to magnetic multilayers (Tsoi et al., 1998; Tsoi et al., 2000). Theory (Slonczewski, 1996; Berger, 1996) connects these features in multilayers with transfer of spin moment between the magnetic layers mediated by polarized current of sufficiently high density.

## 2. Experimental technique

We have measured IVC's, their first $dV/dI(V)$ and the second $d^2V/dI^2(V)$ derivatives for heterocontacts between normal metals (Cu, Ag) and ferromagnetic Co films or bulk ferromagnetic Co, Ni, Fe. 10 or 100 nm thick Co films were deposited onto 100 nm thick bottom Cu electrodes prepared on oxidized silicon substrates. The top Co surface was contacted mechanically by a Cu or Ag needle at low temperature. External magnetic field was applied both perpendicular and parallel to the film plane. The majority of the experiments were carried out on a polycrystalline bulk Co in the form of a parallelepiped 1x1.5x5 mm, using a sharpened silver wire (0.15 mm in diameter) as the needle. Before the measurements the Co surface was electro-polished in a mix of a hydrochloric acid and alcohol (1:1) at current densities of about 2.5 A/cm$^2$ and voltage 8 V. The surface of the Co films was not exposed to any processing. All measurements were done at temperature 4.2 K, and contacts were created directly in liquid helium.

Not all contacts showed PC spectra suitable for further study. Apparently, this is due to the presence of a layer of natural oxide or other imperfections on the surface. So, for example, lightly touching the Co films by the Ag needle could yield contact resistances $R_0$ of several kΩ, and the elasticity of the thin Ag wire was frequently insufficient to further reduce the PC resistance. In such cases low resistance PCs, $R_0$=1–10Ω, were often obtained by switching on/off one of the devices in the measurement circuit, which caused a breakdown of the surface insulating layer. The technique of creating the mechanical contact may also cause significant pressure in the contact area, which can influence the ferromagnetic or-der at the surface. Nevertheless, we have obtained hundreds of PC spectra, which showed interesting spectral and magnetic features.

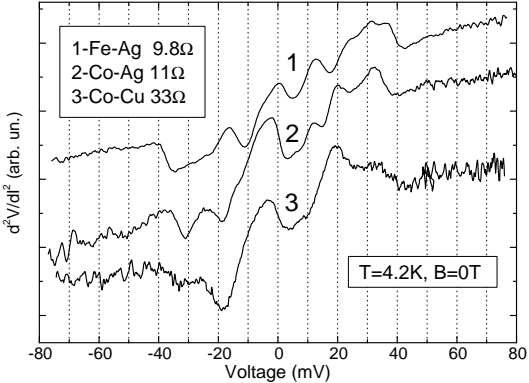

*Figure 1.* Spectra of PCs between bulk Fe or Co and Ag or Cu with distinct phonon maxima of Co or Fe at $|V| > 15$ mV. The Ag phonon maximum around $\pm 12$ mV is seen in curves 1 and 2. The $N$-feature at $V=0$ is typical for magnetic impurities (Naidyuk and Yanson, 2005) (likely Fe or Co atoms) in Ag or Cu (so-called Kondo anomaly).

## 3. Spectra of bulk ferromagnets

Figure 1 shows PC spectra for heterocontacts on bulk Fe and Co in zero magnetic field with extrema symmetrically located relative to $V=0$, which correspond to maxima of the EPI function in Fe or Co and Cu or Ag (see for comparison Fig. 5.9 in (Naidyuk and Yanson, 2005)). The intensity of the phonon maxima varies, which is likely connected with an asymmetrical position of the boundary between the metals relative to the narrowest part of the contact. The spectra with prevailing Co EPI maxima at $V=19$ and $33$ mV (see two bottom curves in Fig. 1) were observed more frequently. Apparently, such a spectrum corresponds to the N/F boundary being close to the PC center, since the intensity of EPI in Co and Fe are higher than those in noble metals (Naidyuk and Yanson, 2005), and the Fermi velocity is lower. It is known (Naidyuk and Yanson, 2005) that in heterocontacts the intensity of the PC spectrum is inversely proportional to the Fermi velocity. Occasionally spectra with clear contributions from the noble metals were observed (see two upper curves in Fig. 1), likely corresponding to deeper penetration of the respective noble metal into the ferromagnet. Thus using the EPI spectrum it is possible to infer the microstructure of the PCs under study.

Figure 2 shows the PC spectrum of another contact between bulk Co and Cu. A maximum in $dV/dI$ is seen at negative bias, which often observed both in zero external field and in magnetic field. This increase in resistance, appearing in $d^2V/dI^2$ as N-shaped feature, can reach several percent of the PC resistance. The N-shaped feature is superposed on the PC EPI spectrum with very smoothed phonon maxima, which indicate a reduction of the electron mean free path. The latter conclusion is based on numerous experimental studies (Lysykh et al., 1980; Naidyuk et al., 1984), where a degradation of phonon maxima was observed upon adding impurities in the contact region, such as 3–10% of Ni in Cu (Lysykh et al., 1980) and 1–3% of Be in Ni (Naidyuk et al., 1984). Thus the probability of observing unipolar N-shaped features increases for PCs with degraded phonon maxima. That is a transition from the ballistic to diffusive and even thermal transport regime favors the appearance of these features.

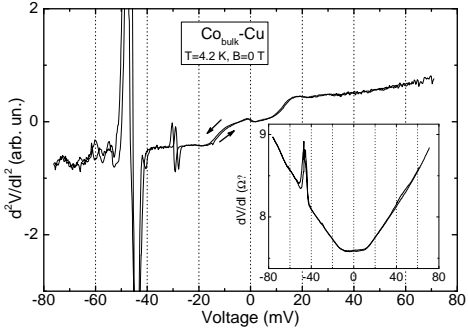

*Figure 2.* $dV/dI$ (inset) and $d^2V/dI^2$ (main panel) for PC bulk Co–Cu, $R_0$=7.5Ω, $B$=0, $T$=4.2 K. Compare the phonon features in $d^2V/dI^2$ at about 20 mV with the maxima in the two bottom curves of Fig. 1. The weak phonon peaks indicate a deviation from the spectral (ballistic) regime of current flow. The maximum in $dV/dI$ at negative bias (inset) corresponds to an N-shaped feature in $d^2V/dI^2$ and is connected with a magnetic excitation in the ferromagnetic layer (see text). $dV/dI$ and $d^2V/dI^2$ were recorded by sweeping the current forth and back, which resulted in a slight horizontal shift due to a relatively large time constant of the lock-in amplifier.

The last is illustrated by Fig. 3, where $d^2V/dI^2$ are presented for PCs with nearly equal resistance, but their spectra have pronounced or smeared phonon maxima with respect to the background at high bias. This illustrates a very typical behavior we observe, namely that the phonon features are suppressed when the N-shaped features are present. A more detailed discussion of this behavior and the underlying physics is given elsewhere (Fisun et al., 2004; Yanson et al., 2005). We did not observe spectra containing simultaneously N-shaped features, illustrated in Fig. 2, and pronounced phonon maxima, shown in Fig. 1. One must note, that not all non-ballistic contacts showed unipolar N-shaped peculiarities. This means that the regime of current flow is a pre-condition for the observed phenomena, which we connect (Fisun et al., 2004; Yanson et al., 2005) with spin torque effects. However, the specific micromagnetic structure in the contact area, affected by possible dislocations, impurities, domain boundaries, etc., must play a role.

A study of Ni–Ag heterocontacts (Fig. 4) revealed, that the N-shaped features in the second derivative of the IVC appear with equal probability in both polarities. The magnetic field behavior is similar, i. e. the field shifts the peaks to higher bias. In Fig. 4(b) spectra with N-shaped features for positive bias polarity, corresponding to an electron flow from ferromagnetic metal into N are shown. Our observation is that such magnetic features appear at positive bias for heterocontacts of Co or Fe with Cu or Ag, however with much lower probability (not more that 10% of spectra having N-anomalies). Moreover, their behavior lacked systematics: they appeared suddenly and unexpectedly disappeared under small changes in contact resistance or in an external magnetic field. We also note, that the probability of occurrence of "magnetic" features increased after polishing the polycrystal Ni with a fine-grained emery paper. Nevertheless, such processing still allowed observation of Ni phonon maxima [see Fig. 4(a)]. We will not dwell further on the nature of these irregular effects, which are likely connected with the specific microscopic structure of the contact(s).

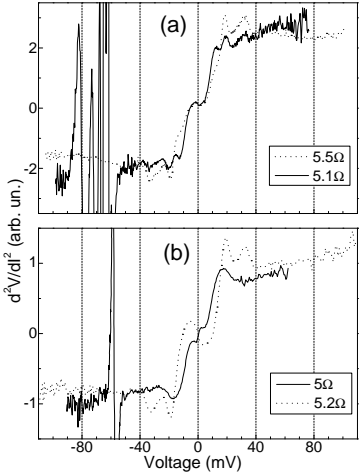

Figure 3. A comparison of the PC spectra for pairs of contacts of nearly equal resistance having different current flow regimes: (a) spectra between bulk Co and Ag; (b) spectra between a 100 nm Co film and Ag (solid line, $B=3$T) or Cu (dotted line, $B=4$T). The threshold current density for the magnetic peaks is about $5 \cdot 10^9$ A/cm$^2$.

## 4. Spectra of film samples

We first present data for 100 nm thick Co films and argue that PCs in this case should not deviate in behavior from those for a bulk Co sample, since the film is thicker than all characteristic length scales in Co – the exchange and spin diffusion lengths. This is because the exchange length in Co is of the order of several nanometers and the spin-lattice relaxation is several tens of nanometers.

Figure 5 shows $d^2V/dI^2$ for a contact between a 100-nm Co film and Cu. Well resolved Co phonon peaks indicate that this contact is close to the ballistic regime. The contribution of Cu in the spectrum, estimated from low intensity of 18-mV peak in comparison with a maximum at 30 mV, is relatively weak. A N-shaped anomaly at $V=0$ in $d^2V/dI^2$ is likely caused by the Kondo effect due to Co atoms penetrating into Cu (Naidyuk and Yanson, 2005). It is interesting to track the behavior of noise at high bias. In Fig. 5 magnetic field of 4 T leads to a reduction of chaotic noise at large negative bias. For more diffusive contacts application of a magnetic field reduces such magnetic noise at the same time often causing N-shaped features at negative bias voltages. In other words, a magnetic field higher than the saturation field for Co ($B_s \approx 1.7$ T) can freeze the spins in one direction, reducing the noise connected with spin reorientation in the PC region.

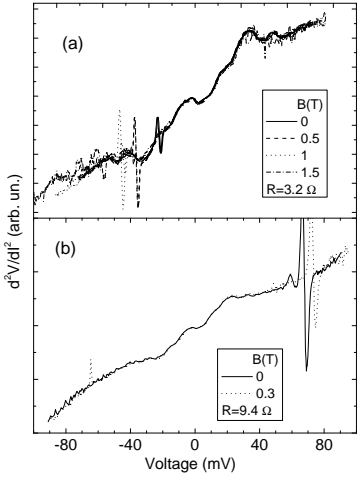

*Figure 4.* $d^2V/dI^2$ of Ni–Ag contacts in a magnetic field applied along needle (saturation field for Ni is 0.6 T). The magnetic N-peculiarities are seen both at negative (a) and positive (b) bias. Magnetic field shifts these features to higher voltage (current) in both cases.

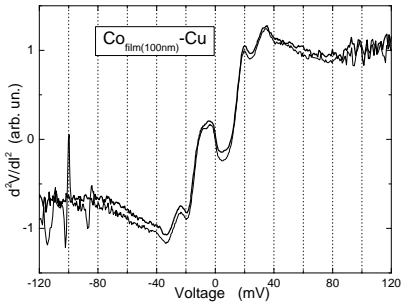

*Figure 5.* $d^2V/dI^2$ spectra of PC between a 100 nm Co film and Cu showing phonon maxima in Co and a Kondo-anomaly at $V=0$ in zero external magnetic field (thin line) and at 4 T (thick line). $R_0 = 2.2\,\Omega$.

The position of the N-shaped features, found at negative bias for a PC with bulk Co as mentioned above, is shifted by an external magnetic field. Such behavior is clearly seen in the PC spectra of Co(10 nm)–Cu contact (Fig. 6). Here the N-shaped features in $d^2V/dI^2$ are observed at negative polarity in fields 0.86–3 T. Their position is shifted proportional to the magnitude of the field [Fig. 6(b)]. At low magnetic fields, the position of the N-shaped feature is unaffected and its intensity falls to nearly zero. Because of the small thickness of the Co film, phonon peaks of the the underlying Cu are seen at 18 and 30 mV. As mentioned above, the first phonon peak of Cu at about 18 mV has a larger intensity than the second peak at about 30 mV, whereas for Co the two phonon peaks are usually comparable in magnitude.

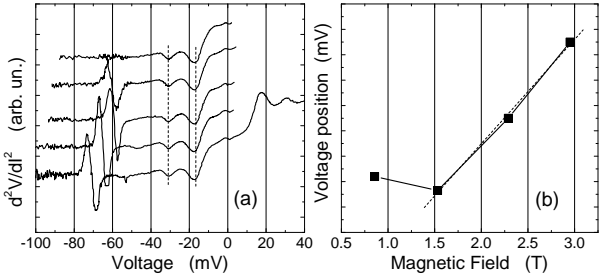

*Figure 6.* (a) Magnetic field dependence of PC spectrum with N-shaped feature for a Co(10 nm)–Cu heterocontact, $R_0$=22 Ω. Magnetic field is: 2.95, 2.29, 1.53, 0.86 and 0 T from bottom. The vertical dashed lines are drawn to mark the positions of the Cu phonon maxima. (b) The position of the N-shaped feature versus magnetic field. Linear approximation of three last points by a dashed line gives a slope $k_I$=0.37 mA/T.

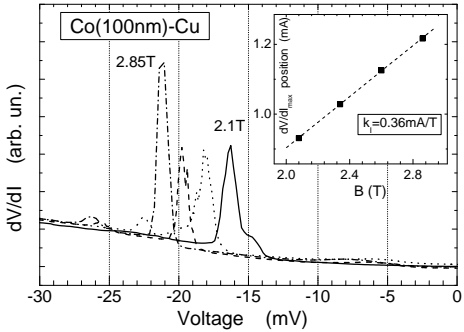

*Figure 7.* Behavior of a peak in $dV/dI$ for a non-spectral PC Co(100 nm)–Cu, $R_0$=17.5 Ω, in magnetic field ($B$=2.85, 2.6, 2.35 and 2.1 T from left to right). Inset: dependence of the peak position versus field. The slope is $k_I$=0.36 mA/T.

In film based contacts, where the PC spectra do not contain phonon peaks or where these are significantly smeared, similar to what is found for PCs to bulk Co, the magnetic peaks in $dV/dI(V)$ appear much more regularly than in contacts with pronounced EPI maxima. For magnetic fields larger than the saturation field, the position of the magnetic peaks on the bias axis is generally linear versus the field [see Fig. 7 and Fig. 6(b)]:

$$I_c \approx I_0 + k_I B \quad or \quad V_c \approx V_0 + k_V B \qquad (1)$$

The transformation in (1) from variable $I$ to $V$ neglects a small ( 10%) non-linearity of the IVC in the given interval of bias. Unipolar maxima in $dV/dI$ have been observed for PCs in (Tsoi et al., 1998; Ji et al., 2003; Rippard et al., 2003) and explained as due to magnetization excitations in the ferromagnet at the interface with the normal metal, caused by a spin-polarized current of high density (about $10^8$ A/cm$^2$). Features observed by us, predominatingly in the non-spectral regime,

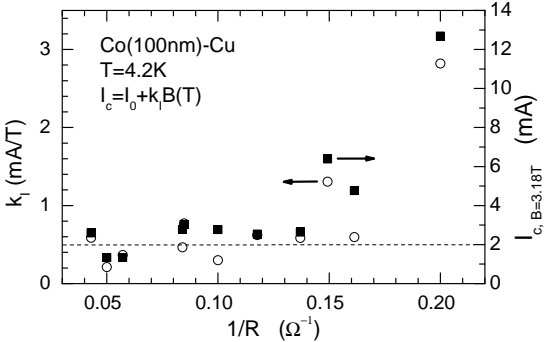

*Figure 8.* Dependence of $k_I$ (open circles) and critical current (solid squares), calculated according to Eq.(1) at $B=3.18$ T, versus inverse contact resistance $1/R$, proportional to contact diameter $d^\alpha$, where $1 < \alpha < 2$

apparently have the same origin. We would like to point out, that the absence of clear phonon features in the spectra of contacts with magnetic anomalies indicates a degradation of the Fermi-step in the electron distribution function by at about the value of the excess energy e$V$, i. e. indicates a current regime close to thermal. The temperature of the lattice, however, is not necessarily in the thermal limit (Naidyuk and Yanson, 2005). Stronger interactions of electrons with magnons (or others nonphononic excitations) can lead to that other sub-system (for example, the spin sub-system) being more intensively warmed up, leading to a loss of the spectral regime of a current flow.

The theory of spin transfer for alternating non-magnetic–ferromagnetic metal structures (Slonczewski, 1996; Berger, 1996) predicts a linear dependence of the critical current $I_c$ on magnetic field, Eq.(1). This critical current produces a jump-like change in the PC resistance. Thus, the derivative of the critical current with respect to field, $dI_c/dB$, should weakly depend on the microstructure of the contact and is defined only by its area and attenuation of spin waves in the ferromagnet (Slonczewski, 1996; Berger, 1996). In the ballistic regime the conductivity of a contact $1/R$ is proportional to its area $S = \pi d^2/4$ (where, $d$ is the PC diameter). In the diffusive regime, $1/R = d/\rho$. Even though the resistivity of a metal in a PC, $\rho$, can considerably differ from that of the respective bulk material, we have attempted to construct the dependence of $dI_c/dB = k_I$ on conductance in Fig. 8. Despite of the significant scatter of points an increase in $k_I$ with increasing $d$ is clearly seen. At small $d$ $k_I$ has a tendency to saturation. $k_I \approx 0.5$ mA/T for the intermediate $d$ range approximately coincides with the values reported by Rippard et al. (Rippard et al., 2003), where the authors studied exchange-coupled Co–Cu multilayer structures and also observed a significant scatter in $k_I$. The $I_c$ versus $1/R$ dependence is consistent with the above behavior at sufficiently high values of the magnetic field. The black squares in Fig. 8 denote $I_c$ at $B=3.18$ T. On the other hand, our attempts to construct a dependence of the critical bias $V_c$ and the corresponding factor $k_V$ on $1/R$ were unsuccessful. It means that the determining parameter in the phenomenon of the magnetic order excitation by a spin-polarized electron flow is the current rather than the voltage drop on the contact.

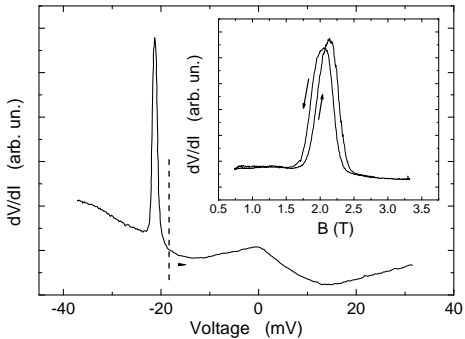

*Figure 9.* A peak in $dV/dI(V)$ at $B=2.96$ T (main panel) transformed into a peak in $dV/dI(B)$ (inset) at bias $V$=-18 mV ($I$=2.3 mA, marked by a vertical dashed line) for a contact between a Co(10 nm) film and Cu, $R = 7.81\,\Omega$.

A step like change in the resistance of N–F PCs is observed not only under action of a high current density, but also under an application of an external magnetic field at a constant bias. In Fig. 9 $dV/dI(V)$ exhibits a magnetic peak at a negative bias of -22 mV. When the bias is fixed at -18 mV and the field is swept in sequence $3.3 \to 0.75 \to 3.3$ T (Fig. 9, inset) a peak is seen in $dV/dI(H)$ at $B$=2.2 T (Fig. 9, inset). The likely origin of this anomaly is that the spins in the contact area form a single domain at high fields which does not contribute additional interface resistance. At low fields, the current from the non-magnetic metal causes a non-uniform spin distribution in the contact, which results in an additional domain wall-like resistance.

## 5. Conclusions

We have analyzed PC spectra of heterocontacts between non-magnetic (normal) metals and ferromagnets, where along with the maxima caused by the electron-phonon interaction, non-spectral features were observed. These features are not connected with the spectral regime of current flow through the contact, where the applied voltage determines the energy of the excitation processes. These non-spectral features have a threshold character, they are observed at some critical values of current through the contacts and connected with excitations of the magnetic subsystem by a current of high density. Such magnetic excitations are observed in diffusive or even more likely thermal contacts, where phonon maxima are smeared or absent due to the electrons experiencing strong impurity scattering. These results demonstrate the importance of the regime of electron flow through a PC for the observed phenomena, namely, that the impurity scattering is found to be at the origin of the new mechanism of single interface spin torque effects. Thus spin torque effects, earlier observed in more complex systems like F–N–F nanopillars and multilayered magnetic structures, are possible for single N–F interfaces provided the transport regime is non-ballistic.

Future research on magnetic point contacts will be directed towards further reduction of the contact size down to dimensions comparable with the ferromagnetic exchange length. PC spectroscopy can be expected to provide additional information on the nature of the spin torque effects in nanometer size magnetic structure, that are beyond the fabrication limits for lithographic techniques.

**Acknowledgements**

The work in Ukraine was partially supported by the National Academy of Sciences within the Program "Nanosystems, nanomaterials and nanotechnologies".